\documentclass[aps,pra,showpacs,tightenlines,twocolumn,lengthcheck]{revtex4-2}
\usepackage{amssymb}
\usepackage{amsmath}
\usepackage{graphicx}
\usepackage{txfonts}
\usepackage{color}

\usepackage{amsfonts}
\usepackage[colorlinks,urlcolor=blue,linkcolor=blue,citecolor=blue]{hyperref}

\begin{document}
	
	\title{Quantum phase transition of the Jaynes-Cummings model}
	\author{Cheng Liu}
	\affiliation{Key Laboratory of Low-Dimensional Quantum Structures and
		Quantum Control of Ministry of Education, Key Laboratory for Matter
		Microstructure and Function of Hunan Province, Department of Physics and
		Synergetic Innovation Center for Quantum Effects and Applications, Institute of Interdisciplinary Studies, Hunan
		Normal University, Changsha 410081, China}
	
	\author{Jin-Feng Huang}
	\email{Corresponding author: jfhuang@hunnu.edu.cn}
	\affiliation{Key Laboratory of Low-Dimensional Quantum Structures and
		Quantum Control of Ministry of Education, Key Laboratory for Matter
		Microstructure and Function of Hunan Province, Department of Physics and
		Synergetic Innovation Center for Quantum Effects and Applications, Institute of Interdisciplinary Studies, Hunan
		Normal University, Changsha 410081, China}
	
	\begin{abstract}
		Herein, we propose an experimentally feasible scheme to show the quantum phase transition of the Jaynes-Cummings (JC) model by modulating the transition frequency of a two-level system in a quantum Rabi model with strong coupling. By tuning the modulation frequency and amplitude, the ratio of the effective coupling strength of the rotating terms to the effective cavity (atomic transition) frequency can enter the deep-strong coupling regime, while the counter-rotating terms can be neglected. Thus, a deep-strong JC model is obtained. The ratio of the coupling strength to resonance frequencies in the deep-strong JC model is two orders of magnitude larger than the corresponding ratio in the original quantum Rabi model. Our scheme can be employed in atom-cavity resonance and off-resonance cases, and it is valid over a broad range. The nonzero average cavity photons of the ground state indicate the emergence of a quantum phase transition. Further, we demonstrate the dependence of the phase diagram on the atom-cavity detuning and modulation parameters. All the parameters used in our scheme are within the reach of current experimental technology. Our scheme provides a new mechanism for investigating the critical phenomena of finite-sized systems without requiring classical field limits, thereby opening a door for studying fundamental quantum phenomena occurring in the ultrastrong and even deep-strong coupling regimes. 
	\end{abstract}
	
	\date{\today}
	\maketitle
	
	\section{Introduction}~\label{introduce}
	The quantum phase transition (QPT)~\cite{sachdev2011}, as an essential concept in physics, is a type of mutation phenomenon and can reveal some important properties of nature. When a nonthermal control parameter, such as coupling strength or a magnetic field, is changed to pass a critical value, the phenomena of spectral gap closing and degeneracy of a ground state with spontaneous symmetry breaking in the system mark the occurrence of a second-order QPT~\cite{Hwang2016}. QPTs have attracted considerable attention and have been extensively studied in various light-matter coupled systems, such as the Dicke model~\cite{Dicke1954,Wang1973,Klaus1973,Huang2009,Baumann2011,Bastidas2012,Klinder2015,Muniz2020,Huang2023}, cavity quantum electrodynamics (QED) lattices~\cite{Greentree2006,Bordyuh2012,Xue2017}, and nonlinear few-body systems~\cite{Felicetti2020}. Usually, a QPT occurs in the thermodynamic limit with a diverging number of components $N\rightarrow\infty$. Recently, a QPT has been observed in finite-sized systems
	in the classical field limit with the ratio $\eta$ for the atomic transition frequency to the cavity frequency tending to infinity, such as the quantum Rabi model~\cite{Ashhab2013,Hwang2015,Chen2021,Cai2021}, finite Jaynes-Cummings (JC) lattices~\cite{Hwang2016}, and Rabi triangle~\cite{Zhang2021,Fallas2022} system. The universal scaling and critical exponent of the anisotropic quantum Rabi model are also discussed by combining the above two limits $N\eta\rightarrow\infty$~\cite{LiuMaoxin2017}. However, these limits put stringent experimental conditions on systems. Therefore, it is natural to ask whether a QPT can occur without the above two limits in a finite-sized system because many important quantum phenomena occur in atom-cavity resonance conditions. Furthermore, external parameters are expected to completely control quantum phases.

	A QPT can occur in the JC model~\cite{Jaynes1963,Huang2020} when the atom-cavity coupling strength enters the deep-strong coupling regime. However, observing this type of QPT in a complete quantum system is challenging because of the non-negligible counter-rotating terms in this coupling regime. When the light is linearly polarized in cavity-QED systems~\cite{Crisp1991} or other platforms that can implement the deep-strong coupling regime of the light-matter interaction, such as the circuit-QED system~\cite{Yoshihara2017} and Landau polaritons~\cite{Bayer2017}, the counter-rotating terms will make important contributions to the system properties. Hence, QPT is difficult to observe in the resonant JC model with atom and cavity resonant coupling.

	Motivated by the above considerations, in this paper, we show QPT in a finite-sized system without the limit $N\eta\rightarrow\infty$. As an example, we study the QPT of the JC model without the classical field limit, i.e., $\eta$ remains finite because the JC model is a fundamental model with finite size in quantum optics.
	We reveal the QPT of the JC model by modulating the transition frequency of a two-level system, while the atom-cavity coupling strength can only be observed in the strong coupling regime. This modulation modifies the cavity frequency and atomic transition frequency. By tuning the modulation frequency and amplitude, the ratio of the effective coupling strength to the effective cavity (atomic transition) frequency can be in a deep-strong coupling regime to obtain an effective deep-strong JC model. Here, this ratio can be increased by two orders of magnitude compared to the original ratio via the modulation. Thus, we can show the rich quantum phases of the JC model. Moreover, we study the average cavity photons of the ground state as an order parameter. The dependence of the phases on the atom-cavity detuning and modulation parameters is also analyzed. The phase diagram guides us to engineer the system via external parameters. Notably, our scheme works far from the ultrastrong coupling regime. All the parameters used in this paper are within the reach of current experimental technology. Our scheme provides a new idea for studying the critical phenomena of finite-sized systems without requiring the classical field limit. Furthermore, the quest for ever greater coupling strength may be slowed down.
	
	\begin{figure}[tbp]
		\center
		\includegraphics[clip, width=8.3cm]{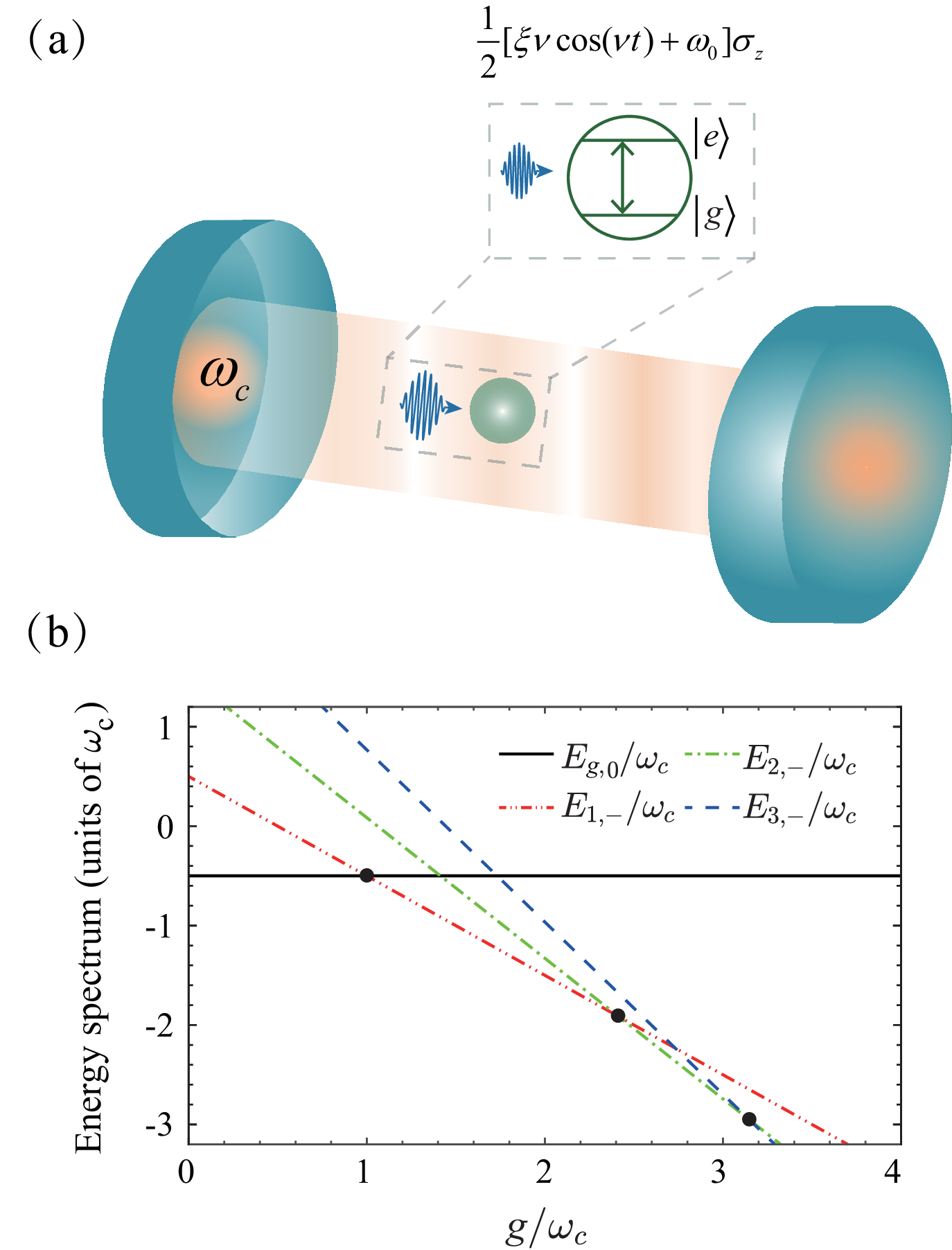}
		\caption{(Color online) (a) Schematic of the system with a two-level system coupled to a single-mode cavity. An applied driving field of the form $\xi\nu\cos(\nu t)\sigma_{z}/2$ modulates the atomic transition frequency.
			(b) Level crossings for the deep-strong Jaynes-Cummings Hamiltonian at resonance $\omega_{0}\!=\!\omega_{c}$. The black dots indicate the transition of the ground states, and their corresponding coupling strengths are $g/\omega_{c}=1$, 2.414, and 3.146.}
		\label{model_v1.eps}
	\end{figure}

	\section{Model and Hamiltonian}~\label{model_hamiltonian}
	We consider a two-level system coupled to a single-mode
	cavity field, described by the quantum Rabi model~\cite{Rabi1936,Rabi1937,Liu20231}, as shown in Figure.~\ref{model_v1.eps}(a). The Hamiltonian of the quantum Rabi model reads ($\hbar =1$)%
	\begin{equation}
		H_{\text{R}}=H_{\text{JC}}+H_{\text{CR}},~\label{Rabi}
	\end{equation}%
	where
	\begin{subequations}
		\begin{align}
			H_{\text{JC}}&=\frac{1}{2}\omega _{0}\sigma _{z}+\omega _{c}a^{\dag
			}a+g( a^{\dag }\sigma _{-}+a\sigma _{+}),\\~\label{JC}
			H_{\text{CR}}&=g(a^{\dagger}\sigma_{+}+a\sigma_{-}).
		\end{align}
	\end{subequations}
	Here, $\sigma _{\pm }=( \sigma _{x}\pm i\sigma _{y}) /2$ are the atomic transition operators with the
	Pauli matrix $\sigma _{x,y,z}$ and transition frequency $\omega
	_{0},$ while $a$ ($a^{\dag }$) is the annihilation (creation) operator of
	the cavity mode with a resonance frequency $\omega _{c}$. We use $\vert e\rangle$ and $\vert g\rangle$ to denote the excited and ground states of the atom, respectively. Further, we have $\sigma_{x}=\vert e\rangle\langle g\vert+\vert g\rangle\langle e\vert$, $\sigma_{y}=-i(\vert e\rangle\langle g\vert-\vert g\rangle\langle e\vert)$, and $\sigma_{z}\!=\!\vert e\rangle\langle e\vert-\vert g\rangle\langle g\vert$. The coupling strength between the atom
	and cavity is $g$. When the coupling strength is far from the ultrastrong coupling regime, i.e., $g/\omega _{c}<0.1$ ($g/\omega
	_{0}<0.1 $), the Hamiltonian $H_{\text{R}}$ can be reduced to the JC
	Hamiltonian under the rotating wave approximation (RWA)~\cite{Jaynes1963}.
	
	The state $\left\vert g,0\right\rangle $ is the ground state of the JC model with an eigenenergy $E_{g,0}=-\omega _{0}/2$ when $g/\omega_{c}<1$ $(g/\omega_{0}<1)$, as indicated by Figure~\ref{model_v1.eps}(b). The other eigenstates are $\vert n,\pm\rangle$ $(n=1,2,\dotsb$) with excitation number $n$, and the corresponding eigenenergy is given by~\cite{Scully1997}
	\begin{equation}
		E_{n,\pm }=\left( n-\frac{1}{2}\right) \omega _{c}\pm \frac{1}{2}\sqrt{%
			4g^{2}n+\delta ^{2}}.~\label{HJC}
	\end{equation}
	Here, $\delta =\omega _{0}-\omega _{c}$ is the detuning between
	the atomic transition frequency and the cavity frequency.
	
	We plot the energy spectrum (units of $\omega_{c}$) of the JC
	Hamiltonian as a function of $g/\omega_{c}$ for $\delta=0$, as shown in Figure~\ref{model_v1.eps}(b). A series of new ground states emerge as $g/\omega_{c}$ enters the deep-strong coupling regime, i.e., $g/\omega_{c}>1$,
	indicating that the QPT occurs in the deep-strong coupling regime. These transition points locate at~\cite{Huang2020}
	\begin{subequations}
		\begin{align}
			g_{0}&=\sqrt{(\omega_{c}+\delta)\omega_{c}},\\
			g_{n}&=\sqrt{\omega_{c}^{2}(2n+1)+\sqrt{4n\omega_{c}^{4}(n+1)+\delta^{2}\omega_{c}^{2}}}
		\end{align}~\label{g0n}
	\end{subequations}
	\\ for $n=1,2,\dotsb$. In particular, the ground state of the system is $\vert 1,-\rangle$ for $1\!<\!g/\omega_{c}\!<\!2.414$ and $\vert 2,-\rangle$ for $2.414\!<\!g/\omega_{c}\!<\!3.146$ at resonance $\delta\!=\!0$. Since the average photon number in the ground state $\vert G\rangle$ is $\langle G\vert a^{\dagger}a\vert G\rangle\!=\!0$ for $g/\omega_{c}\!<\!1$, we label the regime $g/\omega_{c}\!<\!1$ as the normal phase, while if $\langle G\vert a^{\dagger}a\vert G\rangle\!>\!0$ for $g/\omega_{c}\!>\!1$, we label it as the super-radiant phase. The virtual photons in the superradiant phase are bound to the ground state. However, QPT cannot occur in the JC model because of the failure of the
	RWA in the deep-strong coupling regime for finite $\delta$. In the coupling regime $g/\omega_{c}<0.1$, it is promising to neglect the counter-rotating terms. In typical circuit-QED and cavity-QED systems, the coupling strength $g$ is greater than the decay rates $\kappa$ and $\gamma$ of the cavity field and atom, i.e., $g>\kappa, \gamma$. In the following discussion, we focus on the strong coupling regime $\kappa,\gamma<g<0.1\omega_{c},0.1\omega_{0}$ and finite $\delta$.
	
	To reach the effective deep-strong coupling regime, namely, the super-radiant phase of the JC model, we apply a sinusoidal modulation to the two-level system~\cite{Huang2017,Huang2020,Zheng2023}. The corresponding
	Hamiltonian reads
	\begin{equation}
		H_{\text{D}}\left( t\right) =\frac{1}{2}\xi v\cos \left( vt\right) \sigma _{z},
	\end{equation}
	where $v$ and $\xi $ are the modulation frequency and the dimensionless modulation amplitude, respectively. Thus, the total Hamiltonian reads
	\begin{equation}
		H\left( t\right) =\frac{1}{2}\left[ \omega _{0}+\xi v\cos \left( vt\right) \right] \sigma
		_{z}+\omega _{c}a^{\dag }a+g( a^{\dag }+a) \left( \sigma _{+}+\sigma _{-}\right) .
	\end{equation}%
	In the rotating frame defined by the unitary operator%
	\begin{equation}
		U_{1}(t) =\mathcal{T}\exp \left( -i\int_{0}^{t}\left\{\frac{1}{2}\left[ \omega _{0}+\xi v\cos \left( vs\right) \right] \sigma
		_{z}+\omega _{c}a^{\dag }a\right\} ds\right),
	\end{equation}%
	the total Hamiltonian becomes
	\begin{eqnarray}
		H_{\text{rot}}\left( t\right) &=&gJ_{0}\left( \xi \right) e^{i\delta t}a\sigma _{+}+gJ_{0}\left( \xi
		\right) e^{-i\delta t}a^{\dag }\sigma _{-}  \notag \\
		&&+gJ_{m_{0}}\left( \xi \right) e^{i\Delta _{m_0}t}a^{\dag }\sigma
		_{+}+gJ_{m_{0}}\left( \xi \right) e^{-i\Delta _{m_0}t}a\sigma _{-}  \notag \\
		&&+g\sum_{n=-\infty ,n\neq 0}^{\infty }J_{n}\left( \xi \right) \left[
		e^{i\left( \delta +nv\right)t }a\sigma _{+}+e^{-i\left( \delta +nv\right)t
		}a^{\dag }\sigma _{-}\right]   \notag \\
		&&+g\sum_{m=-\infty ,m\neq m_{0}}^{\infty }J_{m}\left( \xi \right)
		\left[ e^{i\Delta _{m}t}a^{\dag }\sigma _{+}+e^{-i\Delta _{m}t}a\sigma _{-}%
		\right]   \label{H_Rt}
	\end{eqnarray}%
	after the
	transformation $U_{1}^{\dagger}\left(t\right)H\left(t\right)U_{1}\left(t\right)+i\dot{U}_{1}^{\dagger}\left(t\right)U_{1}\left(t\right)$. Here, $\mathcal{T}$ denotes the time-ordering operator, $n$ and $m$ are integers, $\Delta
	_{m} =\omega _{0}+\omega _{c}+mv$, and $\Delta _{m_0}$ satisfies
	\begin{equation}
		\vert \Delta _{m_{0}}\vert =\min \left\{ \left\vert \Delta
		_{m}\right\vert =\left\vert \omega _{0}+\omega _{c}+mv\right\vert, \hspace{0.5cm} m\in Z \right\}
	\end{equation}%
	with \textquotedblleft $Z$\textquotedblright\ denoting the set of all integers. In the derivation of $%
	H_{\text{rot}}\left( t\right) ,$ we have used the Jacobi-Anger identity
	\begin{equation}
		e^{iz\sin \zeta }=\sum_{n=-\infty }^{\infty }J_{n}\left( z\right) e^{in\zeta
		}
	\end{equation}
	with the Bessel function of the first kind $J_{n}\left( z\right) $.
	
	We can \textcolor{black}{tune} the modulation frequency $v$ and dimensionless
	modulation amplitude $\xi $ to engineer the system since the oscillating frequency $\delta+nv$ ($\Delta_{m}$) of the sideband for the rotating (counter-rotating) terms and atom-cavity coupling strength $gJ_{n}(\xi)$ depend on $v$ and $\xi$. The sidebands for rotating and counter-rotating terms are separated by $v$. Hence, under the conditions
	\begin{equation}
		v\gg\vert\delta\vert, \hspace{0.5cm}v\gg\vert\Delta_{m_{0}}\vert, \hspace{0.5cm}v\gg g>g\left\vert J_{n}\left( \xi \right) \right\vert ,\label{v_rot}
	\end{equation}
	the terms in the third and fourth lines of eq.~(\ref{H_Rt}) are fast oscillating and can be thus discarded under RWA.
	Then, the Hamiltonian $H_{\text{rot}}\left( t\right) $ can be simplified as
	\begin{equation}
		H_{\textcolor{black}{\text{eff}}}(t) \approx g_{r}(\xi)e^{i\delta t} a\sigma _{+}+g_{c}(\xi)e^{ i\Delta
			_{m_0}t} a^{\dag }\sigma _{+}+\text{H.c.,}  \label{Heff}
	\end{equation}
	
	\begin{figure}[tbp]
		\center
		\includegraphics[clip, width=8.3cm]{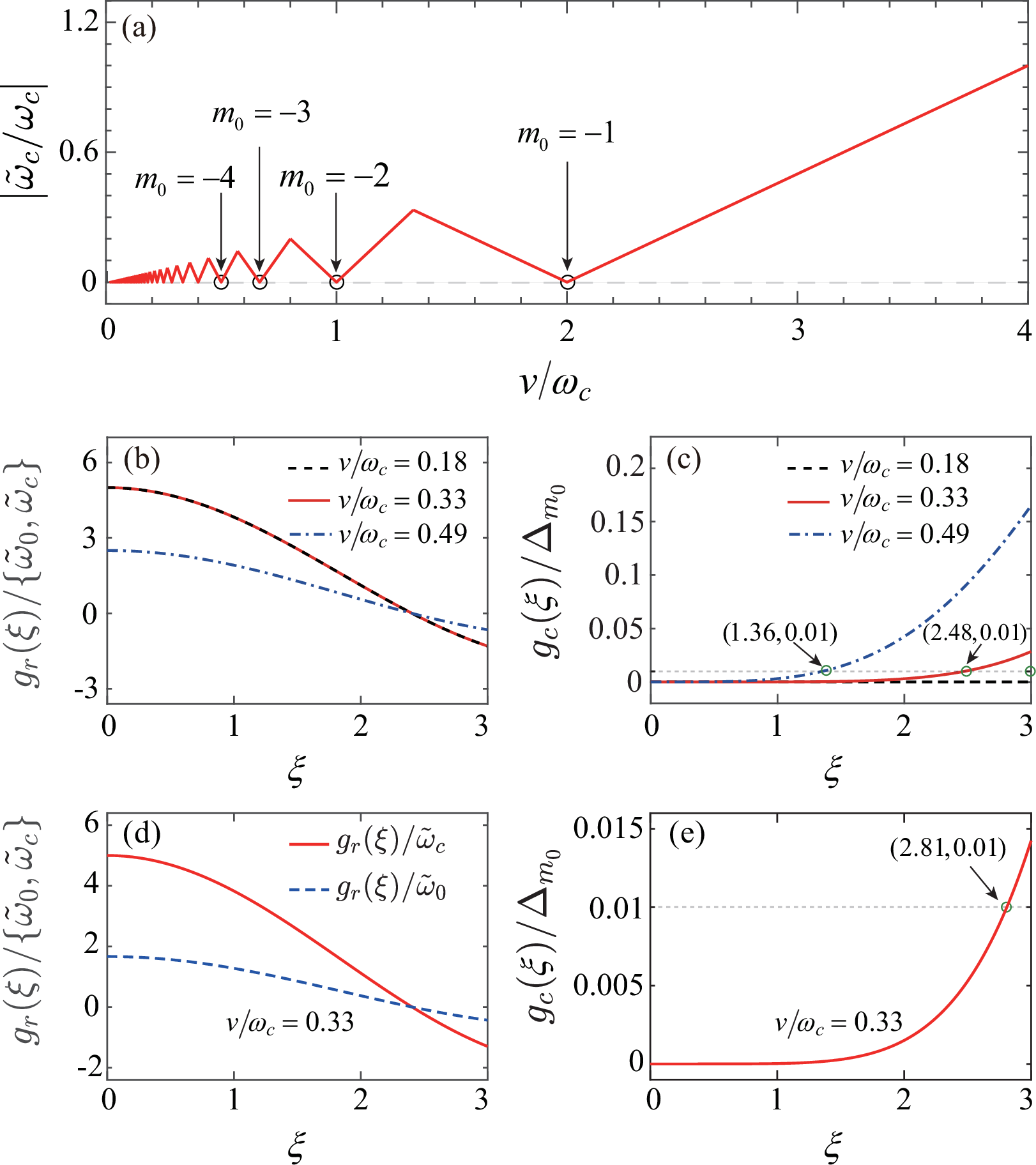}
		\caption{(Color online) (a) Effective frequency of the cavity $\tilde{\omega}_{c}/\omega_{c}$ vs. the modulation frequency $v/\omega_{c}$ at resonance $\delta\!=\!0$.
			(b) and (c): Relative coupling strengths \textcolor{black}{$g_{r}(\xi)/\{\tilde{\omega}_{0},\tilde{\omega}_{c}\}$} and $ g_{c}(\xi)/\Delta_{m_{0}}$ vs. the dimensionless modulation amplitude $\xi$ for $\delta=0$ at various modulation frequencies $v/\omega_{c}\!=\!0.18$ ($m_{0}\!=\!-11$) (black dashed curve), 0.33 ($m_{0}\!=\!-6$) (red solid curve), and 0.49 ($m_{0}\!=\!-4$) (blue dash-dotted curve). (d) and (e): Ratios \textcolor{black}{$g_{r}(\xi)/\{\tilde{\omega}_{0},\tilde{\omega}_{c}\}$} and $ g_{c}(\xi)/\Delta_{m_{0}}$ as a function of $\xi$ for off-resonance case $\delta/\omega_{c}=0.02$ at $v/\omega_{c}=0.33$ ($m_{0}\!=\!-6$). The one sweep at $g_{c}(\xi)/\Delta_{m0}=0.01$ in panels (c) and (e) is indicated by a gray dotted line, and the intersection points are located at $\xi=1.36$ and 2.48 in panel (c) and $\xi=2.81$ in panel (e). The common parameter is $g/\omega_{c}=0.05$. }
		\label{fig2_JC_deep_rwa}
	\end{figure}
	
	\noindent where we have introduced the effective coupling strengths%
	\begin{equation}
		g_{r}(\xi)=gJ_{0}\left( \xi \right) ,\text{ \ }g_{c}(\xi)=gJ_{m_{0}}\left( \xi \right).
	\end{equation}%
	
	Transferring to a second rotating frame by the transformation $U_{2}^{\dagger}\left(t\right)H_{\text{eff}}\left(t\right)U_{2}\left(t\right)+i\dot{U}_{2}^{\dagger}\left(t\right)U_{2}\left(t\right)$ with $U_{2}(t)=\exp[i(\tilde{\omega}_{c}a^{\dagger}a+\tilde{\omega}_{0}\sigma_{z}/2)t]$, the Hamiltonian (\ref{Heff}) becomes
	\begin{equation}
		\tilde{H}_{\text{eff}}=\frac{\tilde{\omega}_{0}}{2}\sigma _{z}+\tilde{\omega}_{c}a^{\dag
		}a+g_{r}(\xi)( a^{\dag }\sigma _{-}+a\sigma _{+}) +g_{c}(\xi)( a^{\dag
		}\sigma _{+}+a\sigma _{-})~\label{H_t_Rabi}
	\end{equation}%
	\textcolor{black}{with the effective frequencies of the atom and the cavity field
		\begin{equation}
			\tilde{\omega}_{0} =\frac{\Delta _{m_{0}}+\delta }{2},\hspace{0.5cm}\tilde{\omega}_{c} =\frac{\Delta _{m_{0}}-\delta }{2},
		\end{equation}
		respectively.} The Hamiltonian $\tilde{H}_{\text{eff}}$ describes an anisotropic Rabi model with the tunable effective frequencies $\tilde{\omega}_{0}(v)$ and $\tilde{\omega}_{c}(v)$ and the effective coupling strengths $g_{r}(\xi)$ and $g_{c}(\xi)$. To obtain an effective deep-strong JC model, we need to set the modulation parameters to satisfy
	\begin{equation}
		g_{c}(\xi)/\Delta_{m_{0}}\ll 1~\label{gcxi}
	\end{equation}
	
	\begin{figure}[tbp]
		\center
		\includegraphics[clip, width=8.3cm]{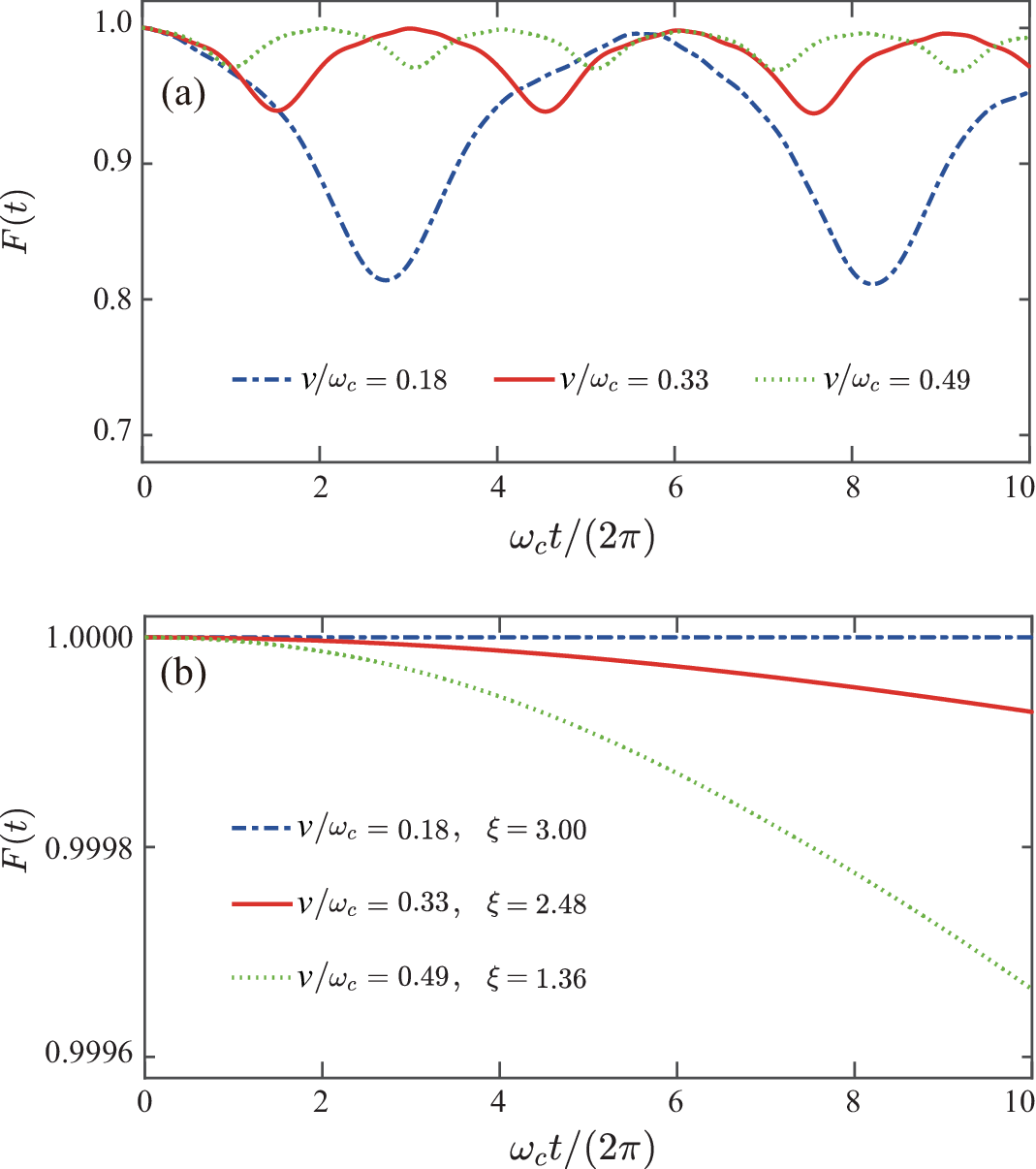}
		\caption{(Color online) Dynamics of the fidelity $F(t)$ in the first (a) and second (b) rotating frames. The fidelity in (a) is obtained using Hamiltonian (\ref{H_Rt}) and (\ref{Heff}), while (b) is obtained utilizing Hamiltonian(\ref{H_t_Rabi}) and (\ref{H_eff_JC}). The initial state of the system is $(\vert g\rangle+\vert e\rangle)\vert \alpha\rangle/\sqrt{2}$ \textcolor{black}{with $\alpha\!=\!0.1$}. In panel (a), we take the modulation amplitude as $\xi=2.48$. The common parameters are $g/\omega_{c}=0.05$ and $\delta=0$.}
		\label{fig5_JC_fidelity}
	\end{figure}
	
	\noindent to make the RWA on eq.~(\ref{H_t_Rabi}).
	
	Under the condition~(\ref{gcxi}), the effective Hamiltonian~(\ref{H_t_Rabi}) can be further simplified as an effective JC Hamiltonian
	\begin{equation}
		\tilde{H}_{\text{JC}}=\frac{\tilde{\omega}_{0}}{2}\sigma _{z}+\tilde{\omega}%
		_{c}a^{\dag }a+g_{r}(\xi)( a^{\dag }\sigma _{-}+a\sigma _{+}) . ~\label{H_eff_JC}
	\end{equation}
	The Hamiltonian describes a two-level system coupled to a single-model cavity field with vacuum Rabi frequency $g_{r}(\xi)$, which depends on the modulation parameter $\xi$ in the form of Bessel function $J_{0}(\xi)$. The effective free frequencies $\tilde{\omega}_{0}(v)$ and $\tilde{\omega}_{c}(v)$ can be tuned using modulation frequency $v$. These dependence relations of \textcolor{black}{\{$\tilde{\omega}_{0}$, $\tilde{\omega}_{c}$\}} and $g_{r}(\xi)$ on ($\xi, v$) guide us on how to reach the deep-strong coupling regime.
	
	\section{Deep-strong coupling}~\label{deepstrongcoupling}
	To reach the deep-strong coupling regime of the effective JC Hamiltonian (\ref{H_eff_JC}), we analyze the validity of the modulation parameters in this section.
	We plot the effective cavity (atomic transition) frequency $\tilde{\omega}_{c}$ ($\tilde{\omega}_{0}$) and coupling strengths $g_{r}(\xi)$ and $g_{c}(\xi)$ as a function of the modulation parameter $v$ or $\xi$ by choosing $g/\omega_{c}\!=\!0.05$, as shown in Figure~\ref{fig2_JC_deep_rwa}. In particular, in Figure~\ref{fig2_JC_deep_rwa}(a), we plot $\tilde{\omega}_{c}$ versus $v$ at resonance $\delta=0$. Note that $\tilde{\omega}_{0}=\tilde{\omega}_{c}$ for atom-cavity resonance at $\delta=0$. A series of V-shaped valleys are found for $v/\omega_{c}<4$. Each V-shaped valley shares the same $m_{0}$ and contains a zero point, which locates at
	\begin{equation}
		v=-\frac{\omega_{0}+\omega_{c}}{m_{0}}.
	\end{equation}
	It is just the \textcolor{black}{observation} that \textcolor{black}{near} the zero point \textcolor{black}{$\tilde{\omega}_{0}=0$ $(\tilde{\omega}_{c}=0)$}, we could possibly make \textcolor{black}{$g_{r}(\xi)/\{\tilde{\omega}_{0},\tilde{\omega}_{c}\}\!>\!1$} by choosing appropriate $v$ and $\xi$. To clearly see the dependence of the effective coupling strengths $g_{r}(\xi)$ and $g_{c}(\xi)$ on $\xi$, in Figures~\ref{fig2_JC_deep_rwa}(b, c),
	we plot \textcolor{black}{$g_{r}(\xi)/\{\tilde{\omega}_{0},\tilde{\omega}_{c}\}$} and $g_{c}(\xi)/\Delta_{m_{0}}$ vs. $\xi$, respectively, for three values of $v$ at $\delta\!=\!0$. As indicated by Figures~\ref{fig2_JC_deep_rwa}(b, c), as $\xi$ increases, \textcolor{black}{$g_{r}(\xi)/\{\tilde{\omega}_{0},\tilde{\omega}_{c}\}$} decreases while $g_{c}(\xi)/\Delta_{m_{0}}$ increases. In particular, at $v/\omega_{c}\!=\!0.18$ ($m_{0}\!=\!-11$) or $0.33$ ($m_{0}\!=\!-6$), \textcolor{black}{$g_{r}(\xi)/\{\tilde{\omega}_{0},\tilde{\omega}_{c}\}$} decreases from 5 to -1.3, while $g_{c}(\xi)/\Delta_{m_{0}}$ remains smaller than $0.01$ for $0\!<\xi\!<3$ and $0\!<\xi<\!2.48$ at $v/\omega_{c}\!=\!0.18$ and $v/\omega_{c}\!=\!0.33$, respectively [Figure~\ref{fig2_JC_deep_rwa}(c)]. \textcolor{black}{Note that $g_{r}(\xi)/\tilde{\omega}_{c}$ remains the same for $v/\omega_{c}=0.18$ and 0.33 since these ratios have the same value of $\tilde{\omega}_{c}$.} For a slightly larger $v$, such as $v/\omega_{c}\!=\!0.49$, \textcolor{black}{$g_{r}(\xi)/\{\tilde{\omega}_{0},\tilde{\omega}_{c}\}$} decreases from 2.5 to -0.64 while $g_{c}(\xi)/\Delta_{m_{0}}<0.01$ for $0\!<\!\xi\!<\!1.36$ [Figure~\ref{fig2_JC_deep_rwa}(c)]. Thus, the effective JC model can enter the deep-strong coupling regime, where the counter-rotating terms are safely neglected by properly choosing $\xi$ and $v$. The above analysis is based on resonance condition $\delta=0$. To further show the validity of eq.~(\ref{H_eff_JC}) in off-resonance case $\delta\neq 0$, we also plot \textcolor{black}{$g_{r}(\xi)/\{\tilde{\omega}_{0},\tilde{\omega}_{c}\}$} and $g_{c}(\xi)/\Delta_{m_{0}}$ vs. $\xi$ for $\delta/\omega_{c}\!=\!0.02$ at $v/\omega_{c}\!=\!0.33$, as shown in Figures~\ref{fig2_JC_deep_rwa}(d, e). We find that \textcolor{black}{$g_{r}(\xi)/\{\tilde{\omega}_{0},\tilde{\omega}_{c}\}$} and $g_{c}(\xi)/\Delta_{m_{0}}$ show similar feature with $\delta=0$. Furthermore, the relative effective coupling strength \textcolor{black}{$g_{r}(\xi)/\{\tilde{\omega}_{0},\tilde{\omega}_{c}\}$} can enter the deep-strong coupling regime while $g_{c}(\xi)/\Delta_{m_{0}}\!<\!0.01$ for $0\!<\!\xi\!<\!2.81$. Our results show that \textcolor{black}{$g_{r}(\xi)/\{\tilde{\omega}_{0},\tilde{\omega}_{c}\}$} can be adjusted over a broad range and enter the deep-strong coupling regime through modulation, while the counter-rotating terms can be neglected. Moreover, with the help of modulation, the \textcolor{black}{ratio $g_{r}(\xi)/\{\tilde{\omega}_{0},\tilde{\omega}_{c}\}$} can be increased by two orders of magnitude compared to the original \textcolor{black}{ratio $g/\{\omega_{0},\omega_{c}\}$}. This result provides us with the mechanism for observing QPT in the JC model with finite atom-cavity detuning $\delta$.

	To further show the validity of the effective JC Hamiltonian (\ref{H_eff_JC}) and verify the above analysis\textcolor{black}{es}, we calculate the fidelity
	\begin{equation}
		F(t)=\left\vert \langle \psi(t)\vert \varphi(t)\rangle\right\vert^{2}
	\end{equation}
	for the above two cases. Because the effective JC Hamiltonian is obtained in the second rotating frame, we compare the dynamics in the first and second rotating frames. The result is shown in Figure~\ref{fig5_JC_fidelity}, where we assume that the initial state of the system is $(\vert g\rangle+\vert e\rangle)\vert \alpha\rangle/\sqrt{2}$. In Figure~\ref{fig5_JC_fidelity}(a), we show the time dependency of the fidelity in the first rotating frame, namely, the fidelity between states $\vert\varphi(t)\rangle$ and $\vert\psi(t)\rangle$, which are governed by the Hamiltonians $H_{\text{rot}}(t)$ in eq.~(\ref{H_Rt}) and $H_{\text{eff}}(t)$ in eq. (\ref{Heff}), respectively, for the same parameters as those in Figures~\ref{fig2_JC_deep_rwa}(b, c) at $\xi=2.48$. The fidelity can reach 81.41\%, 93.88\%, and 97.12\% at $v/\omega_{c}=0.18$, 0.33, and 0.49, respectively. A higher fidelity can be obtained by increasing $v$, where the condition~(\ref{v_rot}) is satisfied better. This result implies that the Hamiltonian (\ref{Heff}) is valid under the current parameters. In Figure~\ref{fig5_JC_fidelity}(b), we present the fidelity in the second rotating frame, obtained by the Hamiltonians (\ref{H_t_Rabi}) and (\ref{H_eff_JC}). The parameters ($v/\omega_{c}$, $\xi$) are chosen as (0.18, 3), (0.33, 2.48), and (0.49, 1.36), indicated by the circles in Figure~\ref{fig2_JC_deep_rwa}(c), denoting \textcolor{black}{$g_{c}(\xi)/\Delta_{m_{0}}\leq 0.01$}. The fidelities are close to 1. Our results show that the effective deep-strong JC Hamiltonian (\ref{H_eff_JC}) is valid in our approximation under the conditions  (\ref{v_rot}) and (\ref{gcxi}).
	
	Under appropriate dimensionless modulation amplitude $\xi$ and frequency $v$, we obtain an effective deep-strong JC Hamiltonian~(\ref{H_eff_JC}). In the effective deep-strong coupling regime, the JC system undergoes a series of QPTs with an increase in the coupling strength, as shown in Figure~\ref{model_v1.eps}(b). In the following section, we will discuss in detail the phenomena of the QPT in this regime.
	
	\section{Manipulation of quantum phases}~\label{QPTJC}
	In this section, we investigate the \textcolor{black}{QPTs} of the effective JC system. By some variable replacements $\omega_{c}\rightarrow\tilde{\omega}_{c}$, $\omega_{0}\rightarrow\tilde{\omega}_{0}$, and $g\rightarrow g_{r}(\xi)$ in eq.~(\ref{HJC}), we can obtain the eigenvalues of the effective JC Hamiltonian (\ref{H_eff_JC}) as follows:
	\begin{equation}
		\tilde{E}_{n,\pm }=\left( n-\frac{1}{2}\right) \tilde{\omega}_{c}\pm \frac{1}{2}%
		\tilde{\Omega}_{n}( \xi), \hspace{0.5cm} n=1,2,\dotsb
	\end{equation}%
	for the eigenstate $\vert n,\pm\rangle$, where
	\begin{equation}
		\tilde{\Omega}_{n}(\xi) =\sqrt{4ng_{r}^{2}(\xi)+
			\tilde{\delta}^{2}},
	\end{equation}%
	and $\tilde{E}_{g,0}\!=\!-\tilde{\omega}_{0}/2$ for the state $\vert g,0\rangle$.
	Here,$  $ the effective detuning $\tilde{\delta}$ is defined as $\tilde{\delta}\equiv\tilde{\omega}_{0}-\tilde{\omega}_{c}=\delta$. The
	eigenstates with excitation number $n$ of the effective Hamiltonian (\ref{H_eff_JC}) can be written as
	\begin{subequations}
		\begin{align}
			\left\vert n,+\right\rangle &=\cos [\theta _{n}(\xi)]\left\vert e,n-1\right\rangle
			+\sin  [\theta _{n}(\xi)]\left\vert g,n\right\rangle,   \\
			\left\vert n,-\right\rangle &=-\sin [\theta _{n}(\xi)]\left\vert
			e,n-1\right\rangle +\cos  [\theta _{n}(\xi)]\left\vert g,n\right\rangle.
		\end{align}
	\end{subequations}
	\textcolor{black}{Here}, the mixing angle $\theta_{n}(\xi)$ satisfies the relations $\sin [2\theta _{n}(\xi)]\!=\!2g_{r}(\xi)\sqrt{n}/\tilde{\Omega}_{n}\left(
	\xi\right)$ and \textcolor{black}{$\cos  [2\theta _{n}(\xi)]\!=\!\delta/\tilde{
			\Omega}_{n}\left( \xi\right) $}.
	
	According to the above analysis\textcolor{black}{es} in Figures~\ref{model_v1.eps}(b) and \ref{fig2_JC_deep_rwa},
	\begin{figure}[tbp]
		\center
		\includegraphics[clip, width=8.3cm]{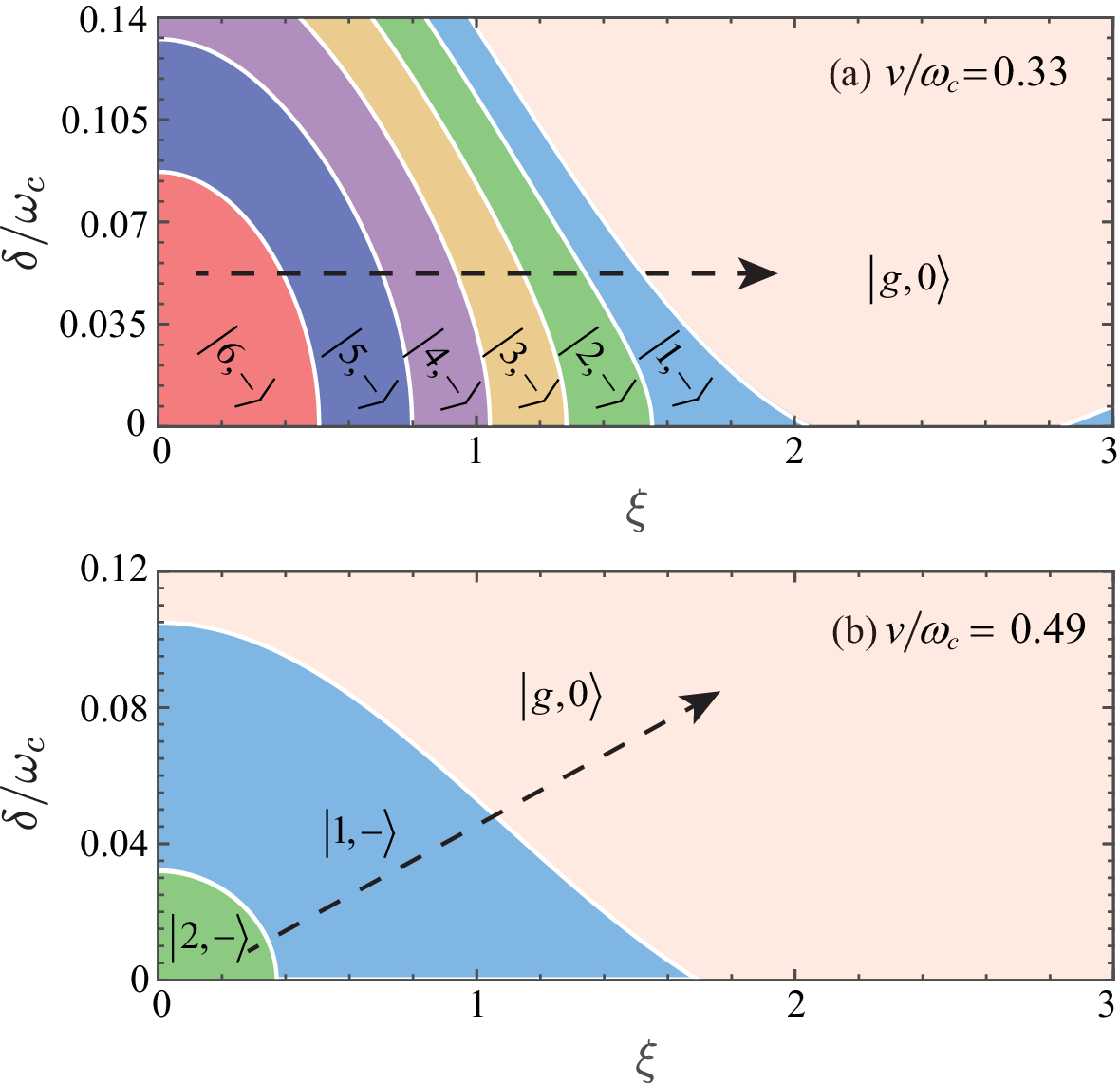}
		\caption{(Color online) Phase diagram vs. the detuning $\delta/\omega_{c}$ and dimensionless modulation amplitude $\xi$ for two modulation frequencies $v/\omega_{c}=0.33$ (a) and 0.49 (b). The other parameter is $g/\omega_{c}=0.05$.}
		\label{fig6_phase_bound_delta_xi}
	\end{figure}
	\noindent when the relative effective coupling strength $g_{r}(\xi)/\tilde{\omega}_{c}$ increases from 0 to the deep-strong coupling regime, i.e., $g_{r}(\xi)/\tilde{\omega}_{c}\geq1$, the ground state of the system varies following the order $\vert g,0\rangle\rightarrow\vert 1,-\rangle\rightarrow\dotsb\rightarrow\vert n,-\rangle$. Similarly, by making the replacements $\omega_{c}\rightarrow\tilde{\omega}_{c}$ in eq.~(\ref{g0n}), the $v$-dependent critical coupling strengths become
	\textcolor{black}{\begin{subequations}
			\begin{align}
				g^{c,0}(v)&=\sqrt{(\tilde{\omega}_{c}+\delta)\tilde{\omega}_{c}},\\
				g^{c,n}(v)&=\sqrt{\tilde{\omega}_{c}^{2}(2n+1)+\sqrt{4n\tilde{\omega}_{c}^{4}(n+1)+\delta^{2}\tilde{\omega}_{c}^{2}}}
			\end{align}~\label{gcn}
	\end{subequations}}\\
	with the photon number $n=1,2,\dotsb$. We can tune the modulation frequency $v$ to vary the critical coupling strengths.
	
	In Figure~\ref{fig6_phase_bound_delta_xi}, we plot the phase diagram against the detuning $\delta/\omega_{c}$ and dimensionless modulation amplitude $\xi$ for two modulation frequencies $v/\omega_{c}\!=\!0.33$ and 0.49. White curves indicate the phase transition boundaries. At $v/\omega_{c}=0.33$, along the arrow indicated in Figure~\ref{fig6_phase_bound_delta_xi}(a), the ground state of the system changes in the following order: $\vert 6,-\rangle\rightarrow\vert 5,-\rangle\rightarrow\vert 4,-\rangle\rightarrow\vert 3,-\rangle\rightarrow\vert 2,-\rangle\rightarrow\vert 1,-\rangle\rightarrow\vert g,0\rangle$ as $\xi$ increases from 0 to 3. For atom-cavity resonance $\delta=0$, along the x-axis, the critical points are located at $\xi=0.505$, 0.797, 1.043, 1.283, 1.553, 2.041, and 2.837. In the regime\textcolor{black}{s} $0\!<\!\xi<\!2.041$ and $2.837\!<\!\xi<\!3$, the system is in the super-radiant phase, while for $2.041\!<\!\xi\!<\!2.837$, it is in the normal phase. Note that without the help of \textcolor{black}{the} modulation when the ground state becomes $\vert 6,-\rangle$,  the atom-cavity coupling strength must reach
	
	\begin{figure}[tbp]
		\center
		\includegraphics[clip, width=8.3cm]{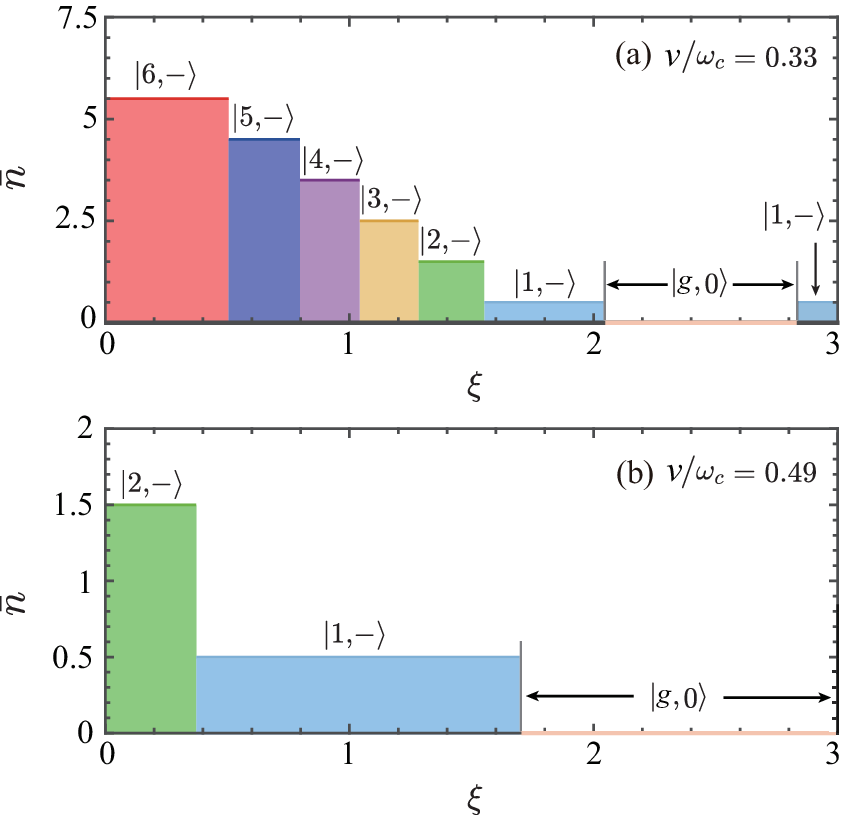}
		\caption{(Color online) Average cavity-photon number $\bar{n}$ in the ground state of the effective Jaynes-Cummings Hamiltonian (\ref{H_eff_JC}) as a function of the dimensionless modulation amplitude $\xi$ for two modulation frequencies $v/\omega_{c}=0.33$ (a) and 0.49 (b). The parameters are $g/\omega_{c}=0.05$ and $\omega_{0}=\omega_{c}$.}
		\label{fig4_JC_photon_ave.eps}
	\end{figure}
	
	\noindent   $g/\omega_{c}=4.686$. Thus, the ground state $\vert 6,-\rangle$ cannot appear without modulation because the counter-rotating terms are important in this deep-strong coupling regime for finite $\delta$. Moreover, to our knowledge, this deep-strong coupling strength
	$g/\omega_{c}=4.686$ has not been implemented experimentally. Here, with the help of the modulation, the atom-cavity coupling strength only requires to be in the strong coupling regime $g/\omega_{c}\!=\!0.05$. To validate our scheme, we also consider another modulation frequency $v/\omega_{c}=0.49$ in Figure~\ref{fig6_phase_bound_delta_xi}(b). The ground state of the system varies in the following order: $\vert 2,-\rangle\rightarrow\vert 1,-\rangle\rightarrow\vert g,0\rangle$ when $\xi$ increases from 0 to 3 along the arrow in Figure~\ref{fig6_phase_bound_delta_xi}(b). In particular, for $\delta=0$, the ground
	state of the system is $\vert 2,-\rangle$ when $0<\!\xi<\!0.372$, while the system is in the normal phase with ground state $\vert g,0\rangle$ when $1.696<\xi<3$.
	\begin{table*}[t]
		\footnotesize
		\textcolor{black}{\caption{Experimental parameters reported in the circuit-QED system~\cite{Chiorescu2004,Johansson2006} and trapped-ion system~\cite{Lv2018}: the resonance frequency $\omega_{0}$ of the two-level system, resonator frequency $\omega_{c}$, coupling strength $g$, and decay rates $\kappa$, $\gamma$ of the resonator and two-level system.}~\label{tab1}}
		\tabcolsep 18pt
		\begin{tabular*}{1\textwidth}{@{\extracolsep{\fill}}c c c  c c  c c c}
			\toprule
			Ref.&Description &$\omega_{0}/2\pi$&$\omega_{c}/2\pi$& $g/2\pi$&$g/\omega_{c}$&$\kappa/2\pi$&$\gamma/2\pi$  \\ \hline
			\cite{Chiorescu2004}& Circuit-QED& 5.9 GHz& 2.91 GHz& 200 MHz&$\sim 0.069$& 1.6 MHz& 27 MHz  \\
			\cite{Johansson2006}& Circuit-QED& 2.1 GHz& 4.35 GHz& 216 MHz&$\sim 0.05$& 0.2 MHz& 0.2 MHz  \\
			\cite{Lv2018}&  Trapped ion& 312.5 kHz& 312.5 kHz& 12.5 kHz&0.04 \\
			\botrule
		\end{tabular*}
	\end{table*}
	\begin{table*}[t]
		\footnotesize
		\textcolor{black}{\caption{Parameters used in our scheme are analyzed using experimental parameters reported in the circuit-QED system~\cite{Johansson2006} and trapped-ion system~\cite{Lv2018}.}~\label{tab2}}
		
		\tabcolsep 8pt
		\begin{tabular*}{1\textwidth}{@{\extracolsep{\fill}}c  c c c  c}
			\toprule
			Notation&Remarks&Used parameters&Circuit-QED&Trapped ion\\ \hline
			$\omega_{c}$&Frequency of the cavity field&Frequency scale&$2\pi\times4.35$ GHz&$2\pi\times312.5$ kHz \\
			$\omega_{0}$&Atomic transition frequency&$\omega_{0}/\omega_{c}=1$&$2\pi\times4.35$ GHz&$2\pi\times312.5$ kHz \\
			$g$& Coupling strength&$g/\omega_{c}=0.05$&$2\pi\times217.5$ MHz&$2\pi\times15.625$ kHz   \\
			$\nu$& Modulation frequency&$\nu/\omega_{c}=0.33$ (0.49)&$2\pi\times1.44$ (2.13) GHz&$2\pi\times103.125$ (153.125) kHz \\
			$\xi$&Dimensionless modulation amplitude&$\xi$=0-3&\\
			$\xi\nu$&Modulation amplitude&$\xi\nu/\omega_{c}=$0$-$0.99 ($0-1.47$)&$2\pi\times$($0-$4.31) [(0$-$6.39)] GHz&$2\pi\times$(0$-$309.375) [($0-$459.375)] kHz  \\
			\botrule
		\end{tabular*}
	\end{table*}
	
	To investigate the ground state of the system in different \textcolor{black}{regimes}, we study the average cavity-photon number of the ground state as the order parameter. For the super-radiant phase with ground state $\vert n,-\rangle$, it is given by
	\begin{equation}
		\bar{n} =\left\langle -,n\right\vert a^{\dag }a\left\vert
		-,n\right\rangle =n-\frac{1}{2}+\frac{\delta }{
			2\tilde{\Omega}_{n}\left(\xi\right) }~\label{nbar}
	\end{equation}%
	and for the normal phase $\bar{n}\equiv\langle g,0\vert a^{\dagger}a\vert g,0\rangle =0$ \textcolor{black}{with the} ground state $\vert g,0\rangle$. Particularly, when $\delta=0$, for the ground state $\vert n,-\rangle$, the average photon number $\bar{n}=n-1/2$ is independent of $\xi$ and $v$. Hence, the nonzero average photon number can indicate the super-radiant phase when the system is in a certain ground state $\vert n,-\rangle$. According to eqs.~(\ref{gcn}) and (\ref{nbar}), the ground state of the system can be controlled by tuning $\xi$ and $v$.
	
	In Figure~\ref{fig4_JC_photon_ave.eps}, we show the average photon number $\bar{n}$ as a function of $\xi$ for
	two modulation frequencies at resonance $\delta=0$. As shown in Figure~\ref{fig4_JC_photon_ave.eps}(a), at $v/\omega_{c}=0.33$, the average photon number decreases in a ladder form as $\xi$ increases from 0 to 2.837, indicating that the ground state varies following the order $\vert 6,-\rangle\rightarrow\vert 5,-\rangle\rightarrow\vert 4,-\rangle\rightarrow\vert 3,-\rangle\rightarrow\vert 2,-\rangle\rightarrow\vert 1,-\rangle\rightarrow\vert g,0\rangle$. All the transition points are consistent with that in Figure~\ref{fig6_phase_bound_delta_xi}(a). In particular, the ground state of the system is in the super-radiant phase when $0\!<\!\xi\!<\!2.041$ and $2.837\!<\!\xi\!<\!3$, corresponding to $1\!<\!g_{r}(\xi)/\tilde{\omega}_{c}\!<\!5$ and $-1.3\!<\!g_{r}(\xi)/\tilde{\omega}_{c}\!<\!-1$, respectively. The normal phase appears when $2.041\!<\!\xi\!<2.837$.
	Similarly, at $v/\omega_{c}=0.49$, as shown in Figure~\ref{fig4_JC_photon_ave.eps}(b), as $\xi$ increases, the ground state varies from $\vert 2,-\rangle$ to $\vert 1,-\rangle$ and then to $\vert g,0\rangle$, sequentially. When $0\!<\!\xi\!<\!1.698$, the ground state of the system is in the super-radiant phase, while its normal phase appears when $1.698\!<\!\xi\!<\!3$. Hence, all phases can be observed here. Thus, the effective JC model can undergo a series of QPTs by tuning the modulation amplitude $\xi$ and frequency $v$.
	
	\section{Discussions on the experimental implementation of parameters }
	\textcolor{black}{In this section, we discuss the experimental parameters used in our scheme. In our scheme,} the super-radiant phase occurs in the JC model when the coupling strength between the atom and cavity field is only in the strong coupling regime with $g/\omega_{c}=0.05$. \textcolor{black}{Various systems can realize this strong coupling regime, such as the circuit-QED system~\cite{Wallraff2004,Chiorescu2004,Johansson2006,Fedorov2010,Li2013,Bosman2017, Forn2019,Blais2021}, trapped-ion system~\cite{Meekhof1996,Leibfried2003,Lv2017,Lv2018,Cai2021}, and cavity-QED system~\cite{Kimble1998}. Recently, even the deep-strong coupling regime $g/\omega_{c}=1.34$ has been reported for the circuit-QED system~\cite{Yoshihara2017}. We show the related parameters reported in the circuit-QED~\cite{Chiorescu2004,Johansson2006} and trapped-ion systems~\cite{Lv2018} in Table~\ref{tab1}. The key point for implementing our scheme is how to manipulate the transition frequency of a two-level system. Currently, the frequency modulation of the atom can be realized in
		the circuit-QED system~\cite{Wallraff2004,Chiorescu2004,Johansson2006,Fedorov2010,Li2013,Bosman2017, Forn2019,Blais2021,Zheng2023}
		and trapped-ion system~\cite{Meekhof1996,Leibfried2003,Lv2017,Lv2018,Cai2021}. Several methods for modulating the frequency of the qubit in the circuit-QED system have been reported~\cite{Johansson2006,Li2013,Forn2019,Zheng2023}. For a superconducting qubit, when applying bias flux through the qubit loop, applying a time-dependent flux can modulate its transition frequency ~\cite{Li2013}. As reported in Ref.~\cite{Li2013}, the cavity frequency and the modulation amplitude are $\omega_{c}/2\pi=3.759$ GHz and $\xi\nu/2\pi=250$ MHz, respectively. The modulation frequency $\nu/2\pi$ can range from 0 to 500 MHz, leading to $\nu/\omega_{c}\in(0, 0.13)$. When $\nu/2\pi=210$ MHz, $\xi\!\approx\! 1.19$. The modulation parameters used in our scheme are of the same order of magnitude as these experimental parameters. Thus, the modulation in our scheme is experimentally accessible in a circuit-QED system.}
	\textcolor{black}{The cavity field can be implemented by an $LC$ resonator~~\cite{Johansson2006}. For an $LC$ resonator with $\omega_{c}/2\pi\!=\!4.35$ GHz, we simulate the parameters in our scheme in Table~\ref{tab2}. As shown in Table 2,}
	these system parameters are experimentally feasible in a circuit-QED system~\cite{Johansson2006,Forn2019}. \textcolor{black}{In a trapped-ion system~\cite{Lv2018},} we can modulate the atomic transition frequency for a two-level system formed by hyperfine states or ions using \textcolor{black}{a pair of counter-propagating pulsed-laser beams with a Raman transition. Similarly, we also simulate parameters used in our scheme using experimental parameters in the trapped-ion system~\cite{Lv2018}}, and they are presented in Table~\ref{tab2}. \textcolor{black}{All parameters in our scheme are experimentally accessible. }
	
	\textcolor{black}{Usually, dissipations affect QPT. Because dissipations can cause the broadening of the ground state energy level and energy level crossing, the critical point is shifted~\cite{Hwang2018,Soriente2018}. In principle, the JC model also experiences QPT in the presence of dissipations. In Figure~\ref{model_v1.eps}(b), we find that the ground state of the system changes from $\vert g,0\rangle$ to $\vert 1,-\rangle$ when the ratio $g/\omega_{c}$ passes the critical point 1. However, we consider the regime in which}
	\textcolor{black}{the decay rates are smaller than $10^{-3}$ of the cavity frequency, i.e., $\{\kappa,\gamma\}/\omega_{c}<10^{-3}$, as shown in Table~\ref{tab1}; hence, we can ignore the effect of the dissipations of the cavity field and atom.}
	
	\textcolor{black}{Now, we estimate the maximum ratio of the effective coupling strength to the effective frequency in our scheme using the parameters reported in Ref.~\cite{Li2013}. According to Ref.~\cite{Li2013}, as discussed above, $\nu/\omega_{c}\in (0,0.13)$ and $\xi$ can reach 1.19. In our scheme, as shown in Figure
		~\ref{fig2_JC_deep_rwa}(b), the ratio $g_{r}(\xi)/\tilde{\omega}_{c}$ increases as $\xi$ decreases, and this ratio reaches its maximum value when $\xi$ approaches 0. For $\nu/\omega_{c}\!=\!0.18$ (or 0.33), which is very close to the experimental parameter in Ref.~\cite{Li2013}, the possible maximum value of $g_{r}(\xi)/\tilde{\omega}_{c}$ is 5 when $\xi\!\rightarrow\! 0$, while the counter-rotating terms can be neglected [Figure~\ref{fig2_JC_deep_rwa}(c)]. Thus, the two orders of magnitude increase in the ratio $g_{r}(\xi)/\tilde{\omega}_{c}$ is experimentally accessible in a contemporary circuit-QED system.}
	\section{Conclusions} \label{disscuss}
	Herein, we proposed a scheme for generating an effective deep-strong JC model by controlling an external classical field. By applying a suitable atomic frequency modulation to a quantum Rabi model in the strong coupling regime, \textcolor{black}{the ratio of the effective coupling strength of the rotating terms to the effective resonance frequency can enter the deep-strong coupling regime}, while the counter-rotating terms can be neglected. \textcolor{black}{This ratio} can be increased by two orders of magnitude through modulation. Thus, we can show the rich quantum phases of the JC model. We calculated the average cavity-photon number as the order parameter for the JC model. The dependence of the quantum phases on the atom-cavity detuning and modulation parameters is obtained. This result provides us with a method for tuning the quantum phases via external modulation parameters. Our work may pave the way for investigating the critical phenomena of finite-sized systems without requiring the classical field limit. Moreover, the quest for ever greater coupling strengths can be slowed down because the quantum phenomena appear\textcolor{black}{ing} in the deep-strong coupling regime may also \textcolor{black}{be} implemented in the strong coupling regime. \textcolor{black}{Most recently, we noticed that the QPT can occur in an open quantum Rabi model in the full quantum limit}~\cite{Filippis2023}.
	
\begin{acknowledgements}
We would like to thank Jie-Qiao Liao for helpful discussions during the reply to the referees' reports. J.-F.H. is supported in part by the National Natural Science Foundation of China (Grant No. 12075083).
\end{acknowledgements}

\end{document}